\documentclass[a4paper,11pt]{article}
\usepackage{pos}
\usepackage{ulem}
\usepackage{graphicx}
\usepackage{subfig}
\usepackage{gensymb}
\usepackage{floatrow}
\newfloatcommand{capbtabbox}{table}[][\FBwidth]

\title{SST-1M Observations of Markarian 421}
\author*[a]{S.~R.~Muthyala}
\author[a]{A.~Araudo}
\author[a]{J.~Jury\v{s}ek}
\author[a]{A.~L.~Müller}

\affiliation[a]{\textit{FZU - Institute of Physics of the Czech Academy of Sciences,\\ Na Slovance 1999/2, Prague 8, Czech Republic}}

\onbehalf{for the SST-1M Collaboration}

\emailAdd{muthyala@fzu.cz}

\abstract{Markarian 421 (Mrk 421) is the closest and one of the brightest high-frequency peaked blazars, located at a redshift of z = 0.031. It is a strong source of gamma rays, and its broadband emission has been extensively studied over the years through multi-wavelength observations from various telescopes.
Mrk 421 has been a target of observational campaigns conducted by the SST-1M telescopes – two single-mirror small-size Cherenkov telescopes at Ondrejov Observatory, Prague, Czech Republic. These telescopes operate in mono and stereoscopic modes, utilizing the Imaging Atmospheric Cherenkov Technique (IACT) to detect Very High Energy (VHE) gamma rays in the 1–300 TeV energy range.
We present recent SST-1M observations, data analysis, and the results of preliminary physical modeling of Mrk 421's emission mechanisms.}

\ConferenceLogo{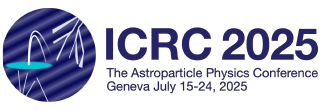}

\FullConference{39th International Cosmic Ray Conference (ICRC2025)\\
 15–24 July 2025\\
Geneva, Switzerland\\}

\begin{document}
\maketitle

\section*{Introduction to SST-1Ms}

Observations of very high energy (VHE) gamma rays can shed light on the highly accelerating
processes in the universe. However, VHE gamma rays cannot be directly detected by ground-based telescopes, as they are absorbed in the atmosphere. Instead, they can be detected by ground-based telescopes through Cherenkov radiation produced by charged particles resulting from the interactions of gamma-ray photons with atomic nuclei in the atmosphere.

\begin{figure}[!h]
  \centering
  \subfloat{\includegraphics[width=0.45\textwidth]{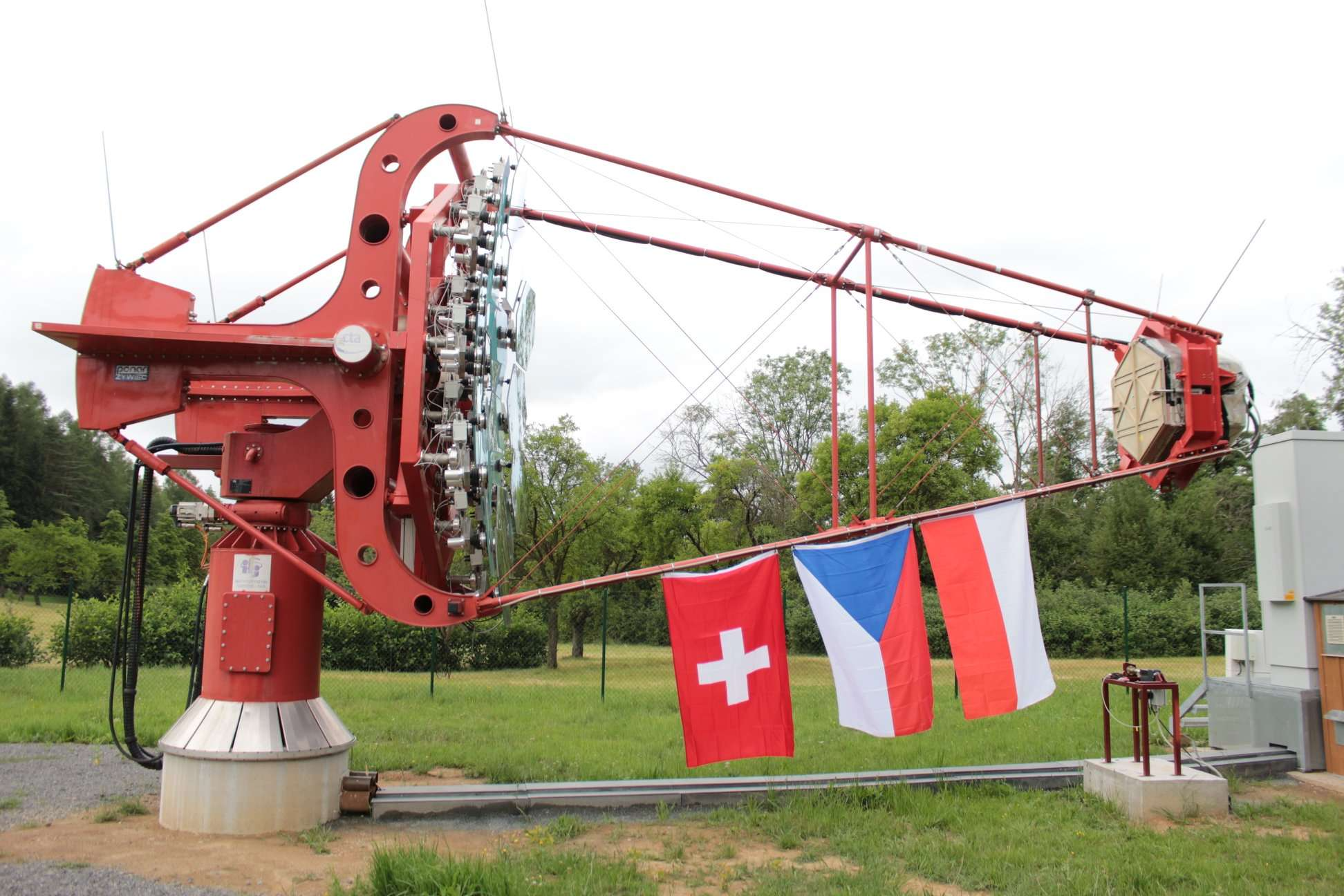}}
   \hfill
  \subfloat{\includegraphics[width=0.55\textwidth]{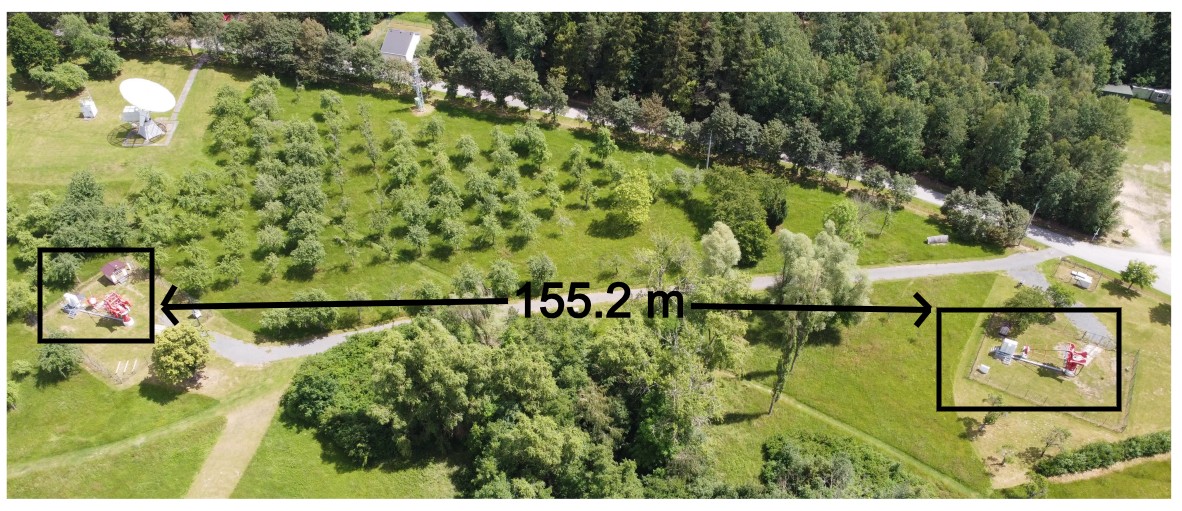}}
  \caption{LEFT: SST-1M telescope 1 in Ondrejov Observatory near Prague. RIGHT: An aerial view of the Ondřejov site shows the two SST-1Ms located 155.2 m apart (C. Alispach et al.,~2025). }\label{fig: fig1}
\end{figure}

SST-1M is a single-mirror small-size Cherenkov telescope as shown in Figure ~\ref{fig: fig1}. Ondrejov observatory in the Czech Republic hosts two SST-1M telescopes, which can operate in mono mode and also take advantage of their relative distance of 155.2~m and detect the showers stereoscopically (see right panel of Figure~\ref{fig: fig1}) \citep{sst1m_hw_paper, sst1m_performance_paper}. The SST-1M timestamps are synchronized to nanosecond precision using the White Rabbit timing network \citep{white_rabbit}. SST-1M was designed by a consortium of institutes from Poland, Switzerland, and the Czech Republic, aiming for the detection of the VHE gamma-rays induced atmospheric showers in the energy range of 3-300 TeV. SST-1M uses a 4-m diameter, single multi-segmented mirror dish composed of 18 hexagonal facets and a highly performing SiPM-based camera with a wide optical field of view of 9 degrees. These SST-1Ms are designed and built to enable remote observations and can be used in mono and stereo modes. The geometrical and timing properties of the detected waveforms are used to find the physical properties of the primary particle. The raw pixel waveforms are calibrated and processed up to the photon list with reconstructed energies and arrival directions (see \citep{sst1m_performance_paper} for the details of the analysis and reconstruction) using the standard data processing and analysis tool developed for SST-1M, \texttt{sst1mpipe}\footnote{\url{https://github.com/SST-1M-collaboration/sst1mpipe}} \citep{sst1mpipe_073}.

\section{Markarian 421}

Blazars are a particular class of radio-loud active galactic nuclei (AGN), characterized by the ultra-relativistic jets from the supermassive black hole being oriented along or very close to the line of sight. Due to relativistic beaming, the jet pointing toward us appears highly boosted and often dominates the observed emission across the entire electromagnetic spectrum, while the counter-jet is typically undetectable. Blazars are the most commonly detected class of VHE gamma-ray sources in the extragalactic sky. Their spectral energy distribution (SED) is characterized by two energy peaks. The low-energy peak is produced by the synchrotron emission from ultra-relativistic electrons embedded in the magnetic field of the jet, while the high-energy peak is thought to be produced by synchrotron self-Compton (SSC) emission \cite{Albert_22}. Based on the position of the synchrotron peak, these sources are subclassified as low synchrotron peaked (LSP) blazars if the peak frequency is less than 10$^{14}$~Hz, or high synchrotron peaked (HSP) blazars if the frequency is larger than 10$^{14}$~Hz.

Mrk 421 is an HSP blazar and a strong gamma-ray source. It is the closest and one of the brightest blazars, at a redshift  $z = 0.031$ or a distance of 122~Mpc from the Earth. It was first detected above 0.5~TeV by the Whipple Observatory in 1992 \cite{1992Natur.358..477P}. This source is continuously monitored by the Large Area Telescope (LAT) onboard NASA's Fermi Gamma-Ray Space Telescope in the energy range of 50~MeV to 1~TeV. Recently, the High Altitude Water Cherenkov (HAWC) Observatory reported photons with energies up to 9 TeV \cite{Albert_22}. It is a highly variable source; hence, several multi-wavelength studies have been performed by many telescopes over the years. 

\section{Data Acquisition and Analysis}

In this contribution, we present the first SST-1M observations of Mrk 421, conducted between
January and May 2024, and derive a preliminary model of emission using the averaged SED. During this observation period, the raw data collected by the telescopes for 23 days of effective exposure in mono mode is about 55.24~h hours for telescope 1 and 60.60~h hours for telescope 2, of which 51.10~h hours are available in stereoscopic mode. Since the atmospheric conditions at Ondřejov are highly variable, in order to attain a high-quality data set, we perform data selection. Data quality cuts are applied to remove data affected by bad atmospheric conditions, such as highly variable night sky background (NSB), clouds, auroras, and technical issues. After the selection cuts, the final sample used for data analysis consists of about 34.31 hours for telescope 1 and 40.48 hours for telescope 2, of which 32.92 hours are available in stereoscopic mode.

For observation and data taking, we follow the wobble technique, which allows us to determine the background from the same data set as the signal events \citep{1994APh.....2..137F}. When the gamma-ray shower triggers the SST-1M telescope, the entire waveform of 50 samples with 4 ns binning is stored. During data analysis in \texttt{sst1mpipe}, the waveform in each camera pixel is calibrated, integrated, and cleaned from background noise to obtain the image parameters. For these cleaned images, Random Forests trained on the Monte Carlo simulations are applied to each event to get the final parameters of the primary particles, including the so-called gammaness, describing how gamma-like a given shower image is. An energy-dependent cut on gammaness is applied to obtain the final photon list. From the final photon list, we can derive the fluxes and energies to construct spectral energy distributions (SEDs) and skymaps.

\begin{figure} 
  \centering
  \subfloat{\includegraphics[width=0.65\textwidth]{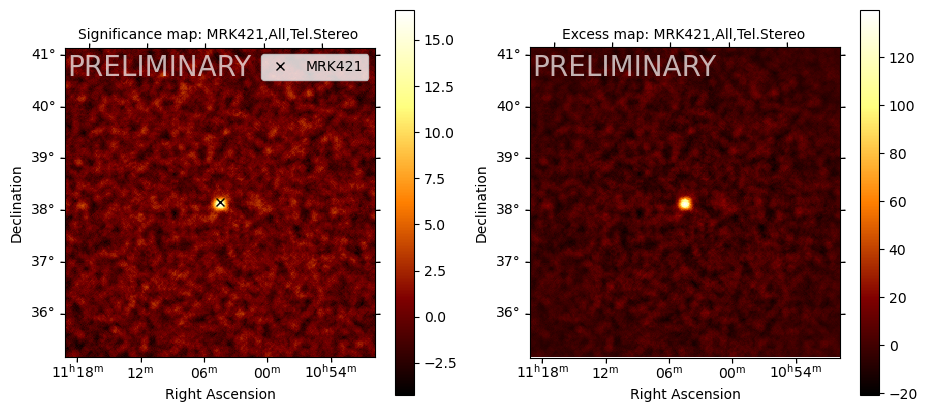}}
   \hfill
  \subfloat{\includegraphics[width=0.35\textwidth]{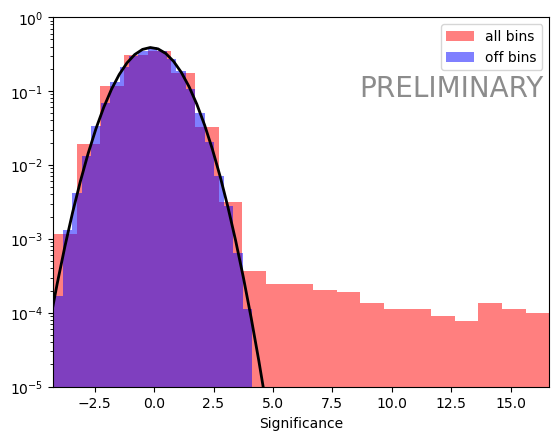}}
  \caption{Significance (LEFT) and excess (MIDDLE) maps of Mrk 421 in stereo mode. RIGHT: 1D distribution of significance in stereo mode.}\label{fig:fig2}
\end{figure}

The significance map in the left panel of Figure~\ref{fig:fig2} quantifies the statistical probability of observing the signal at a particular location in the sky. The excess map in the middle panel of Figure~\ref{fig:fig2} shows the difference between the observed gamma-ray counts and the expected background counts. The maps are produced using the 'ring background method' in the Python package \texttt{gammapy} \citep{axel_donath_2021_5721467}. Here, a ring with radius  $0.7{\degree}$ and width  $0.3\degree$ was used to estimate background events, and a region of $0.3\degree$ around Mrk~421 was excluded from the background estimation. The source was clearly detected during the observations of 32.9 hours with a total significance of 10.95 and excess counts of 178. The distribution of local significance in the right panel of Figure~\ref{fig:fig2} shows that the background/off-region can be described by a Gaussian function, and the excess counts show the presence of the source. The background distribution is fitted by a Gaussian with a mean of $-0.14$ and a standard deviation of $1.03$, which is consistent with theoretical expectations for an unbiased background.

\begin{figure} 
    \centering
    \includegraphics[scale=0.45]{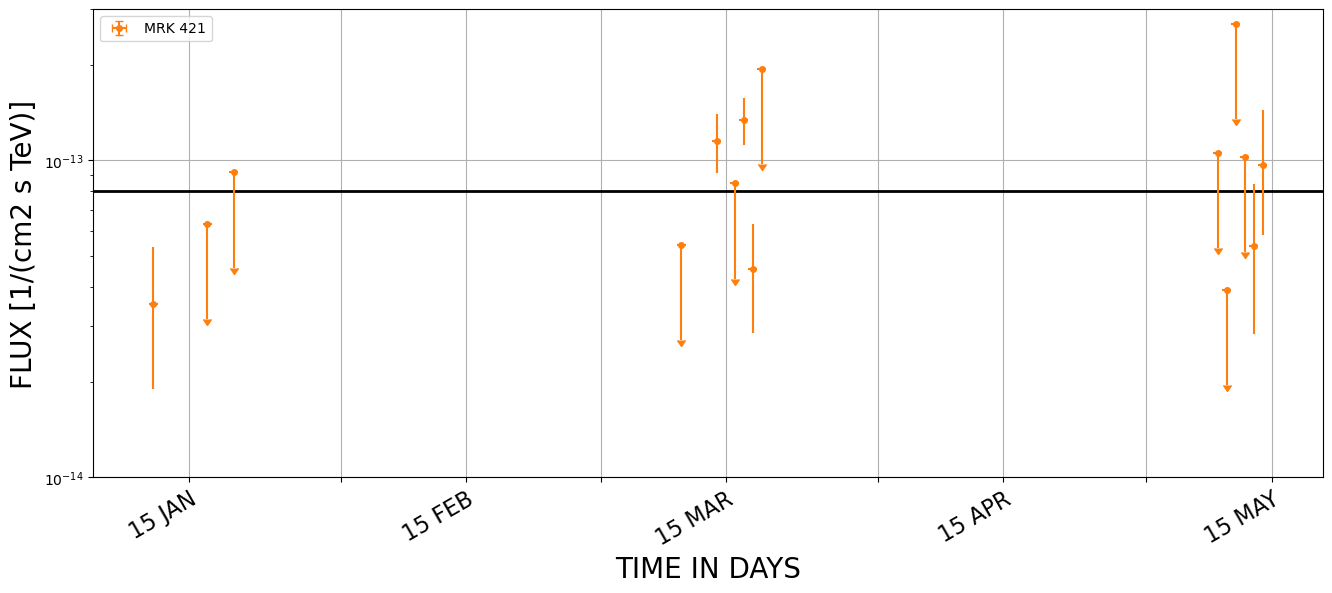}
    \caption{Light Curve of Mrk 421 in stereo mode for the year 2024.}
    \label{fig:fig3}
\end{figure}

We also used \texttt{gammapy} to perform 1D spectral analysis using the reflected background method. The forward folding method is used to reconstruct the gamma-ray spectrum observed by the telescope. This is one of the methods used to remove the experimental systematics. It aims at finding the true distribution by maximizing the agreement probability between the experimental data distribution and the one expected from Monte Carlo. For every night, we reconstruct the spectrum using the Power Law (PL) spectral model ${\rm d}N/{\rm d}E = A (E/3\,{\rm TeV})^{-\alpha}$, where $\alpha$ stands for spectral index and $A$ for the flux normalization at 3 TeV. 
In Figure~\ref{fig:fig3} we show the light curve of Mrk~421, where the total flux per night is obtained by integrating $\int ({\rm d}N/{\rm d}E) {\rm d}E$. If we assume a constant flux from Mrk421, we obtain an average flux for energies above 1 TeV of $F = [8.01 \pm 2.54] \times 10 ^{-14} $ (Tev$^{-1}$cm$^{-2}$s$^{-1}$). In the figure, flux data points are represented by orange markers and the black line represents the average flux. Note that no flares are detected in this observation period \footnote{Typically, for a flare, the peak flux should be more than three times the standard deviation of the average flux in a band.} (see however, \cite{ATel}). 

\begin{figure} 
    \centering
    \includegraphics[scale=0.55]{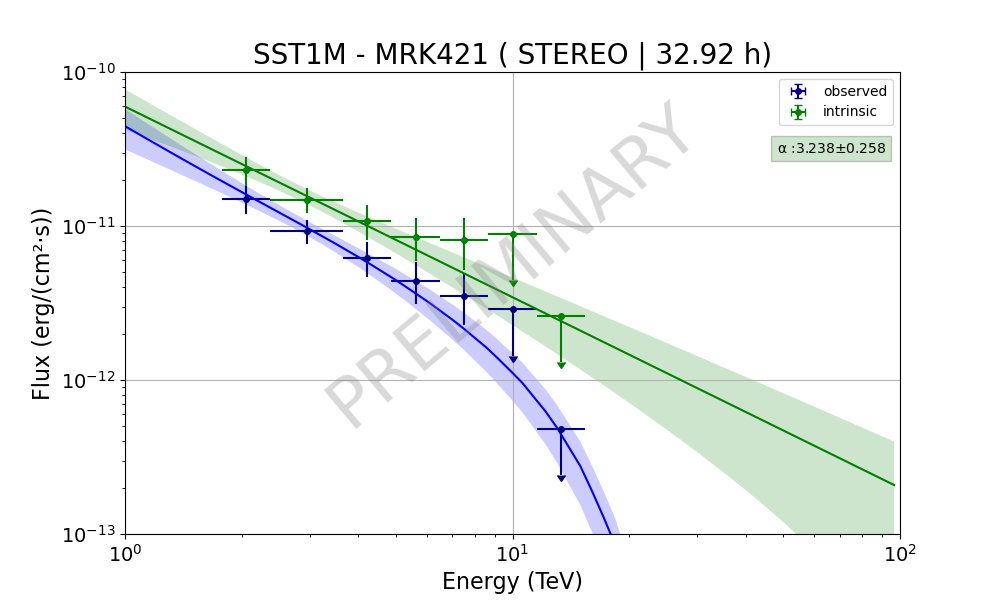}
    \caption{Energy spectrum of Mrk~421 in stereo mode, showing both observed (blue) and intrinsic (green) spectra fitted with a PL model.}
    \label{fig:fig4}
\end{figure}

The preliminary spectrum of Mrk 421 obtained from the final data files of the source in stereo mode using roughly 32.92~h of observations is shown in Figure~\ref{fig:fig4}. The image shows the observed and intrinsic spectrum produced using a PL spectral model in blue and green, respectively, along with the uncertainty bands of the model (area), the derived flux data points, and upper limits in the same colors.  The best-fit parameters of intrinsic spectra are shown in Table~\ref{table:1}.

\begin{table} 
\centering
\begin{tabular}{|c|c|c|c|}
 \hline
  Best fit parameters & Power law index & Amplitude (Tev$^{-1}$cm$^{-2}$s$^{-1}$) & Reference Energy (TeV) \\ \hline
  Intrinsic Spectrum  & $3.24 \pm 0.26$ & $[ 1.06 \pm 0.12 ] \times 10^{-12}$ & $3.00 \pm 0.00$  \\
 \hline
\end{tabular}
\caption{Best fit parameters for the intrinsic  spectra of Mrk~421 obtained using PL model.}
\label{table:1}
\end{table}

\section{Data Modeling}

Previous studies \citep[e.g.][]{Albert_22} show a good fit for the low and high-energy components, but a poor fit for the high-energy tail, usually due to a lack of sufficient data. In this study, we aim to obtain better fits of the high-energy tail of Mrk~421 with the addition of new data observed by SST-1M at photon energies greater than 2~TeV.


\begin{figure} 
  \centering
  \subfloat{\includegraphics[width=0.65\textwidth]{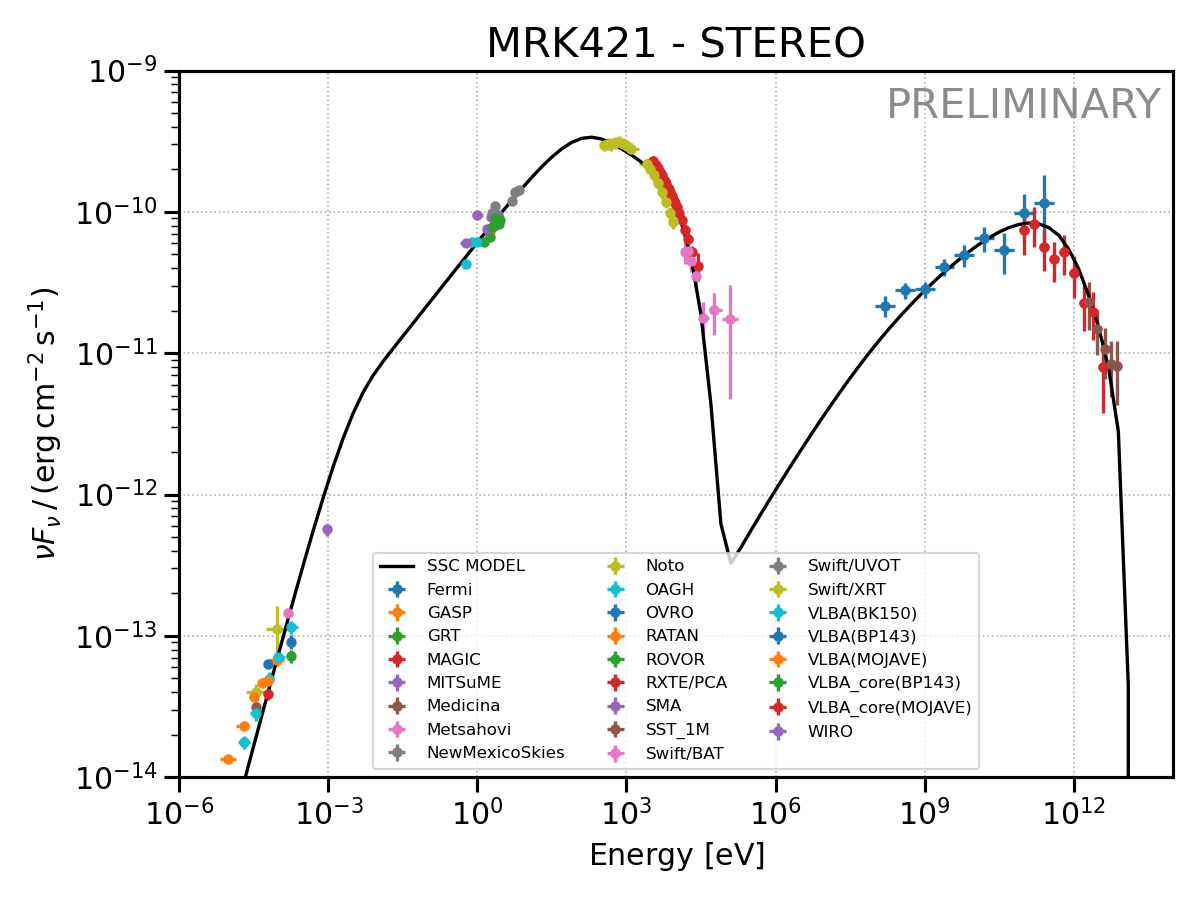}}
   \hfill
  \subfloat{\includegraphics[width=0.35\textwidth]{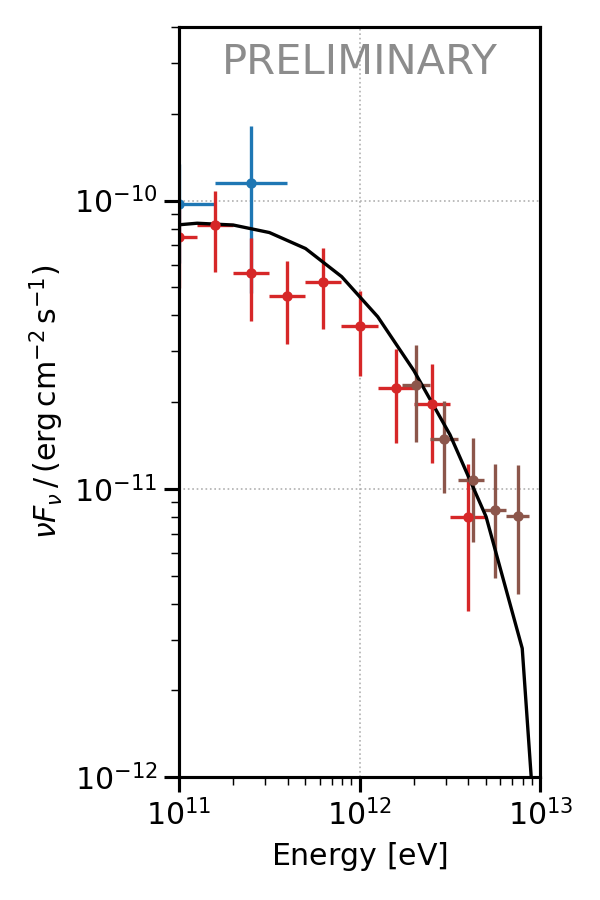}}
  \caption{LEFT: SED of Mrk 421 with best fit SSC model using agnpy package. RIGHT: zoomed figure showing SST-1M data points on the SED.}\label{fig: fig5}
\end{figure}

 Following \cite{Albert_22}, we use the Python package \texttt{agnpy} \citep{2022A&A...660A..18N}, and a $\chi^2$ fit is then performed by the Python package \texttt{gammapy}. We consider a broken power law for the distribution of relativistic electrons with a minimum energy of $500 m_e c^2$. The data consists of observed data from SST-1M combined with the time-averaged SED of Mrk~421 from Abdo et al (2011a) \citep{Abdo_2011}, which provides multi-frequency data from several instruments covering the energy range from radio to gamma rays, observed during the time period of January 19 to June 1, 2009. In their study, the data were corrected for host galaxy emission subtraction, galactic extinction correction for optical/x-ray data, and TeV data were corrected for absorption in Extragalactic Background Light (EBL) using the model in \cite{Franceschini_17}. The best fit corresponds to a Doppler factor of  $\delta_{D}$ = $ 24.29 \pm 0.01 $, giving the value of radius emission zone as $R = 6.1 \times10^{16}$~cm using values of redshift as $z = 0.031$ and the variability time scale of 1 day; and for an electron distribution that follows a broken power law with energy break of $E_{\rm b} = 59.08 \pm 0.51$~GeV and spectral indexes before and after the break of $\alpha_1 = 2.11 \pm 0.01 $ and $\alpha_2 = 3.38 \pm 0.10 $, respectively. The minimum electron energy was frozen to $E_{\rm min} = 255$~MeV, and the maximum electron energy of the best fit is given by $E_{\rm max} = 557.47 \pm 0.58$~GeV. The magnetic field results in a value of B = $ 24.58 \pm 1.01$~mG. In Figure~\ref{fig: fig5} we show the multi-frequency SED of Mrk421 (data points) and the best SSC model (black line), and a summary of the best-fit parameters is shown in Table 2, which are in good agreement with previous results \cite{Albert_22}.

\begin{table} 
\centering
\begin{tabular}{|c|c|c|}
 \hline
 Parameter & Symbol & Mrk 421 \\ \hline
 Doppler factor & $\delta_{D}$ & $ 24.29 \pm 0.01 $ \\ 
 Magnetic field & B [mG]& $ 24.58 \pm 1.01 $ \\ 
 Spectral index before break & $\alpha_1$ &  $ 2.11 \pm 0.01 $ \\ 
 Spectral index after break & $\alpha_2$ & $ 3.38 \pm 0.10 $ \\ 
 Energy break & $E_{\rm b}$ [GeV] &  $ 59.08 \pm 0.51 $ \\  
 Maximum electron energy & $E_{\rm max}$ [GeV] & $ 557.47 \pm 0.58 $ \\ 
 \hline
\end{tabular}
\caption{Best fit parameters for the SSC model of Mrk 421 using agnpy.}
\label{table:2}
\end{table}

\section{Conclusions}

We present the preliminary results of the first set of observations of Mrk~421 taken by SST-1M telescopes in stereo mode from January to May 2024, with roughly 33 hours of observation. We obtain an intrinsic energy spectrum of Mrk~421 by fitting a PL model with an index of $3.24 \pm 0.26$ and an amplitude of $[ 1.06 \pm 0.12 ] \times 10^{-12}$~TeV$^{-1}$~cm$^{-2}$~s$^{-1}$, at a reference energy of 3~TeV. Significance and excess maps show the detection confidence of the source. The best-fit parameters of the SED modeling of Mrk~421 in stereo mode seem to be in good agreement with previous results \cite{Albert_22}. In a further study, we will focus on improving the SED by adding data from HAWC \cite{Albert_22}, and more simultaneous data from other instruments at different energies to build a time-averaged SED model of Mrk~421.

\section{ Acknowledgments }
This publication was created as part of the projects funded in Poland by the Minister of Science based on agreements number 2024/WK/03 and DIR/\-WK/2017/12. The construction, calibration, software control, and support for operation of the SST-1M cameras are supported by SNF (grants CRSII2\_141877, 20FL21\_154221, CRSII2\_160830, \_166913, 200021-231799), by the Boninchi Foundation, and by the Université de Genève, Faculté de Sciences, Département de Physique Nucléaire et Corpusculaire. The Czech partner institutions acknowledge support of the infrastructure and research projects by the Ministry of Education, Youth and Sports of the Czech Republic (MEYS) and the European Union funds (EU), MEYS LM2023047, EU/MEYS CZ.02.01.01/00/22\_008/0004632, CZ.02.01.01/00/22\_010/0008598, Co-funded by the European Union (Physics for Future – Grant Agreement No. 101081515), and Czech Science Foundation, GACR 23-05827S.


\bibliographystyle{JHEP}
\bibliography{Mrk}

\clearpage
\section*{Full Authors List: SST-1M Collaboration}
\scriptsize
\noindent
C.~Alispach$^1$,
A.~Araudo$^2$,
M.~Balbo$^1$,
V.~Beshley$^3$,
J.~Bla\v{z}ek$^2$,
J.~Borkowski$^4$,
S.~Boula$^5$,
T.~Bulik$^6$,
F.~Cadoux$^`$,
S.~Casanova$^5$,
A.~Christov$^2$,
J.~Chudoba$^2$,
L.~Chytka$^7$,
P.~\v{C}echvala$^2$,
P.~D\v{e}dic$^2$,
D.~della Volpe$^1$,
Y.~Favre$^1$,
M.~Garczarczyk$^8$,
L.~Gibaud$^9$,
T.~Gieras$^5$,
E.~G{\l}owacki$^9$,
P.~Hamal$^7$,
M.~Heller$^1$,
M.~Hrabovsk\'y$^7$,
P.~Jane\v{c}ek$^2$,
M.~Jel\'inek$^{10}$,
V.~J\'ilek$^7$,
J.~Jury\v{s}ek$^2$,
V.~Karas$^{11}$,
B.~Lacave$^1$,
E.~Lyard$^{12}$,
E.~Mach$^5$,
D.~Mand\'at$^2$,
W.~Marek$^5$,
S.~Michal$^7$,
J.~Micha{\l}owski$^5$,
M.~Miro\'n$^9$,
R.~Moderski$^4$,
T.~Montaruli$^1$,
A.~Muraczewski$^4$,
S.~R.~Muthyala$^2$,
A.~L.~Müller$^2$,
A.~Nagai$^1$,
K.~Nalewajski$^5$,
D.~Neise$^{13}$,
J.~Niemiec$^5$,
M.~Niko{\l}ajuk$^9$,
V.~Novotn\'y$^{2,14}$,
M.~Ostrowski$^{15}$,
M.~Palatka$^2$,
M.~Pech$^2$,
M.~Prouza$^2$,
P.~Schovanek$^2$,
V.~Sliusar$^{12}$,
{\L}.~Stawarz$^{15}$,
R.~Sternberger$^8$,
M.~Stodulska$^1$,
J.~\'{S}wierblewski$^5$,
P.~\'{S}wierk$^5$,
J.~\v{S}trobl$^{10}$,
T.~Tavernier$^2$,
P.~Tr\'avn\'i\v{c}ek$^2$,
I.~Troyano Pujadas$^1$,
J.~V\'icha$^2$,
R.~Walter$^{12}$,
K.~Zi{\c e}tara$^{15}$ \\

\noindent
$^1$D\'epartement de Physique Nucl\'eaire, Facult\'e de Sciences, Universit\'e de Gen\`eve, 24 Quai Ernest Ansermet, CH-1205 Gen\`eve, Switzerland.
$^2$FZU - Institute of Physics of the Czech Academy of Sciences, Na Slovance 1999/2, Prague 8, Czech Republic.
$^3$Pidstryhach Institute for Applied Problems of Mechanics and Mathematics, National Academy of Sciences of Ukraine, 3-b Naukova St., 79060, Lviv, Ukraine.
$^4$Nicolaus Copernicus Astronomical Center, Polish Academy of Sciences, ul. Bartycka 18, 00-716 Warsaw, Poland.
$^5$Institute of Nuclear Physics, Polish Academy of Sciences, PL-31342 Krakow, Poland.
$^6$Astronomical Observatory, University of Warsaw, Al. Ujazdowskie 4, 00-478 Warsaw, Poland.
$^7$Palack\'y University Olomouc, Faculty of Science, 17. listopadu 50, Olomouc, Czech Republic.
$^8$Deutsches Elektronen-Synchrotron (DESY) Platanenallee 6, D-15738 Zeuthen, Germany.
$^9$Faculty of Physics, University of Bia{\l}ystok, ul. K. Cio{\l}kowskiego 1L, 15-245 Bia{\l}ystok, Poland.
$^{10}$Astronomical Institute of the Czech Academy of Sciences, Fri\v{c}ova~298, CZ-25165 Ond\v{r}ejov, Czech Republic.
$^{11}$Astronomical Institute of the Czech Academy of Sciences, Bo\v{c}n\'i~II 1401, CZ-14100 Prague, Czech Republic.
$^{12}$D\'epartement d'Astronomie, Facult\'e de Science, Universit\'e de Gen\`eve, Chemin d'Ecogia 16, CH-1290 Versoix, Switzerland.
$^{13}$ETH Zurich, Institute for Particle Physics and Astrophysics, Otto-Stern-Weg 5, 8093 Zurich, Switzerland.
$^{14}$Institute of Particle and Nuclear Physics, Faculty of Mathematics and Physics, Charles University, V Hole\v sovi\v ck\' ach 2, Prague 8, Czech~Republic.
$^{15}$Astronomical Observatory, Jagiellonian University, ul. Orla 171, 30-244 Krakow, Poland.

\end{document}